\documentclass[a4paper,11pt]{article}
\pdfoutput=1 

\usepackage{jinstpub} 
\usepackage{color,soul}
\usepackage{enumitem}
\usepackage{subfig}
\usepackage{siunitx}
\usepackage{microtype}

\newcommand{\sixhundred}{370}
\newcommand{\twohundred}{124}

\newcommand{\twenty}{12}
\newcommand{\ten}{6}
\newcommand{\six}{3.7}
\newcommand{\fourpointfive}{2.8}
\newcommand{\one}{0.6}

\title{\boldmath Radiation-induced Effects on Data Integrity and Link Stability of the RD53A Pixel Readout Chip}

\author[a,1]{M. Vogt\note{Corresponding author.}}

\affiliation[a]{University of Bonn,\\Nussallee 12, 53115 Bonn, Germany}

\emailAdd{vogt@physik.uni-bonn.de}

\abstract{	
	The phase-2 upgrade of the LHC will require novel pixel readout chips, which deliver hit information at drastically increased data rates and tolerate unprecedented radiation levels.
	The large-scale prototype chip RD53A has been designed by the RD53 collaboration and manufactured in a \SI{65}{nm} CMOS process, suitable for the innermost layers of both the ATLAS and the CMS experiment.
	In order to verify the radiation hardness design goal of \SI{500}{Mrad} total ionizing dose, RD53A has been irradiated using X-rays. The radiation effects on the performance of the data link, reset circuit and the clock generation have been investigated. Furthermore, the operating margins in terms of supply voltage and frequency have been analyzed.
}


\collaboration[c]{on behalf of the RD53 collaboration}

\proceeding{PIXEL 2018, 9$^{\text{th}}$ International Workshop on Semiconductor Pixel Detectors for Particles and Imaging\\
	December 12-14, 2018\\
	Taipei, Taiwan}

\begin{document}
	\maketitle
	\flushbottom
	
	\section{Introduction}
	\label{sec:intro}
	The phase-2 upgrade of the Large Hadron Collider (LHC)~\cite{HL-LHC} will substantially increase the instantaneous luminosity by a factor of at least 5. This requires novel pixel readout chips with highly complex digital architectures, which deliver hit information at drastically increased data rates. Unprecedented radiation tolerance is essential, especially close to the interaction point.
	The RD53 collaboration~\cite{rd53a_website} was formed to approach these challenges by designing a prototype pixel readout chip in a \SI{65}{nm} CMOS technology, which is suitable for the innermost layers of the pixel detector in the ATLAS~\cite{atlas} and CMS~\cite{cms} experiments.
	
	The large scale prototype chip RD53A~\cite{rd53a} has been produced and is available since December 2017. The stability and locking behavior of the high speed Aurora links of non-irradiated RD53A chips have been investigated in lab environments with different cables, powering schemes, Command Data Recovery (CDR) configurations and synchronization patterns. Performance characterization measurements of the output drivers are ongoing.
	The readout system BDAQ53~\cite{bdaq_git} has been developed~\cite{bdaq} to perform characterization- and test beam measurements. It consists of an FPGA-based readout board and a Python-based data acquisition and analysis framework.
	
	In order to understand the degradation of the clocking- and communication periphery of RD53A, dedicated irradiation campaigns are necessary. Various test routines monitored the chip performance during the irradiation.
	Based on the results, methods to improve the operating parameters and design changes for the upcoming RD53B chip submission will be evaluated.

	\section{The RD53A pixel readout chip}
	\label{sec:RD53A}
	RD53A combines three different analog front end designs and two memory architectures, which are currently being reviewed, to select one of each for the second prototype chip RD53B.
	A new design concept was used for the pixel matrix: It is divided into 8x8 pixel cores, which are assembled from 2x2 pixel analog quads, embedded in synthesized digital logic. A pixel core can be simulated on transistor-level. All pixel cores are identical, which allows for efficient hierarchical verification~\cite{veripix}.

	\section{Total Ionizing Dose effects}
	\label{sec:rad}
	One of the most important Total Ionizing Dose (TID) surface damage mechanisms of MOS transistors is charge trapping, which takes place in the thin field oxide between the bulk substrate and the polysilicon layer. Thereby, transistors can alter their threshold voltage and transconductance, which leads to speed degradation and increased jitter in digital circuits.
	Modern \SI{65}{nm} CMOS processes have been evaluated \cite{65nm} and show relatively high intrinsic radiation tolerance \cite{1grad}, which makes them suitable for high radiation environments. However, especially narrow transistors suffer from additional TID effects like increased leakage current.

	\section{TID campaign}
	\label{sec:tid}
	In the inner layers of the new pixel detector for the High Luminosity LHC, a total ionizing dose of up to \SI{1}{Grad} is expected, including a safety factor.
	The pixel detector however will be replaced after \SI{500}{Mrad} and the readout chip is designed to meet this specification. The radiation damage by TID is a cumulative effect and thus, accelerated high dose-rate campaigns are a common method to evaluate the radiation hardness of semiconductor devices. However, lower dose rate exposure, as present in the experiment, may yield different results. Dedicated irradiation studies are inevitable.
	
	The transistor performance in the \SI{65}{nm} process used for RD53A, was investigated up to \SI{1}{Grad} \cite{1grad}. Radiation effects on different digital standard cells in \SI{65}{nm} have been investigated with simulations and measurements using the dedicated test chip DRAD~\cite{drad}. The \SI{500}{Mrad} radiation models developed within RD53 collaboration overestimate the TID damage level, because they assume worst-case bias conditions at room temperature and don't consider annealing effects. Further TID studies with RD53A are necessary to test the simulation predictions.

	\subsection{Setup and methodology}
	\label{sec:setup}
\begin{figure}[htbp]
		\centering
		\includegraphics[width=.9\textwidth,origin=c]{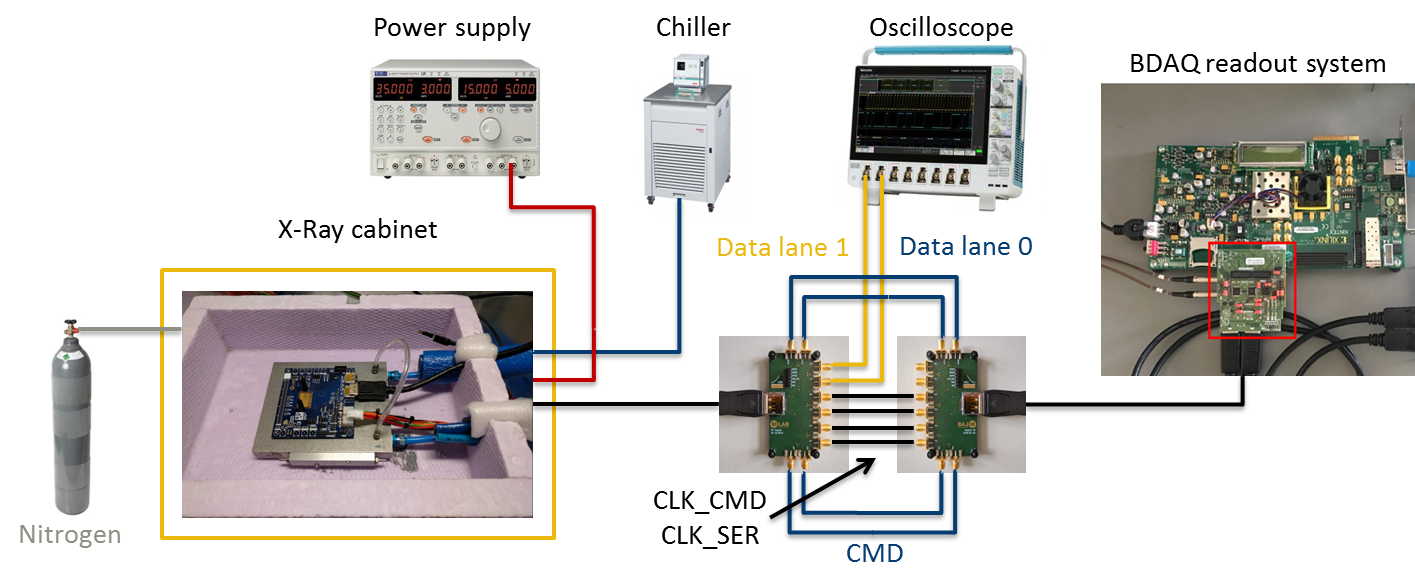}\\[-2ex]
		\caption{\label{fig:setup}X-ray irradiation setup.}
	\end{figure}

	The irradiation campaign was carried out using a new X-ray facility in Bonn (figure \ref{fig:setup}). The tungsten target tube was operated at an acceleration voltage of \SI{40}{\kilo\volt}. During the irradiation, the chip was cooled to \SI{-5}{\degreeCelsius} in dry $\rm N_2$ atmosphere and operated with a monitor script, which included digital scans, temperature and power consumption measurements.
	
	In order to observe effects of charge trapping, it is important to create realistic biasing conditions. To ensure this, all three front ends of the pixel matrix were active and clocked during the irradiation.
	Due to problems with the initial dosimetry, the actual dose was lower than expected. Instead of the anticipated dose of \SI{600}{Mrad}, only \SI{\sim\sixhundred}{Mrad} were achieved.
	This total ionizing dose was reached in multiple steps with dose rates of \SI{\fourpointfive}{Mrad/h} for the first \SI{\twenty}{Mrad} and \SI{\six}{Mrad/h} for the rest of the campaign.
	At each TID step, time consuming and detailed measurements like full shmoo scans (section \ref{sec:shmoo}) and eye diagrams were performed.
	
	The final pixel readout chips will be operated at \SI{1.2}{\volt} and a nominal system clock frequency of \SI{160}{\mega\hertz}. During this campaign, RD53A was operated at an extended range of digital supply voltage, $\rm V_{DDD}=0.8 - 1.3\ \rm V$ and input clock frequency, $\rm f_{CMD}=140 - 180\ \rm MHz$, in order to determine the operating margin before and after irradiation.
	At each combination within this parameter space, tests were performed to evaluate the reliability of operation.
	In order to disentangle the radiation effects on the digital logic and the clock recovery circuit (section~\ref{sec:link}), the chip was operated in the so called CDR-bypass mode. This allows to externally supply two reference clocks, which normally are generated by the chip.
	The readout system BDAQ53 allows to operate RD53A in this mode by providing two variable frequency clocks with fixed phase relation: The previously mentioned $\rm f_{CMD}$ and a second, faster clock $\rm f_{SER}$, for the output data links.

	\subsection{Reference current}
	\label{sec:iref}
\begin{figure}[htbp]
		\centering
		\subfloat[\label{fig:iref:a}]{\includegraphics[width=.46\textwidth,origin=c]{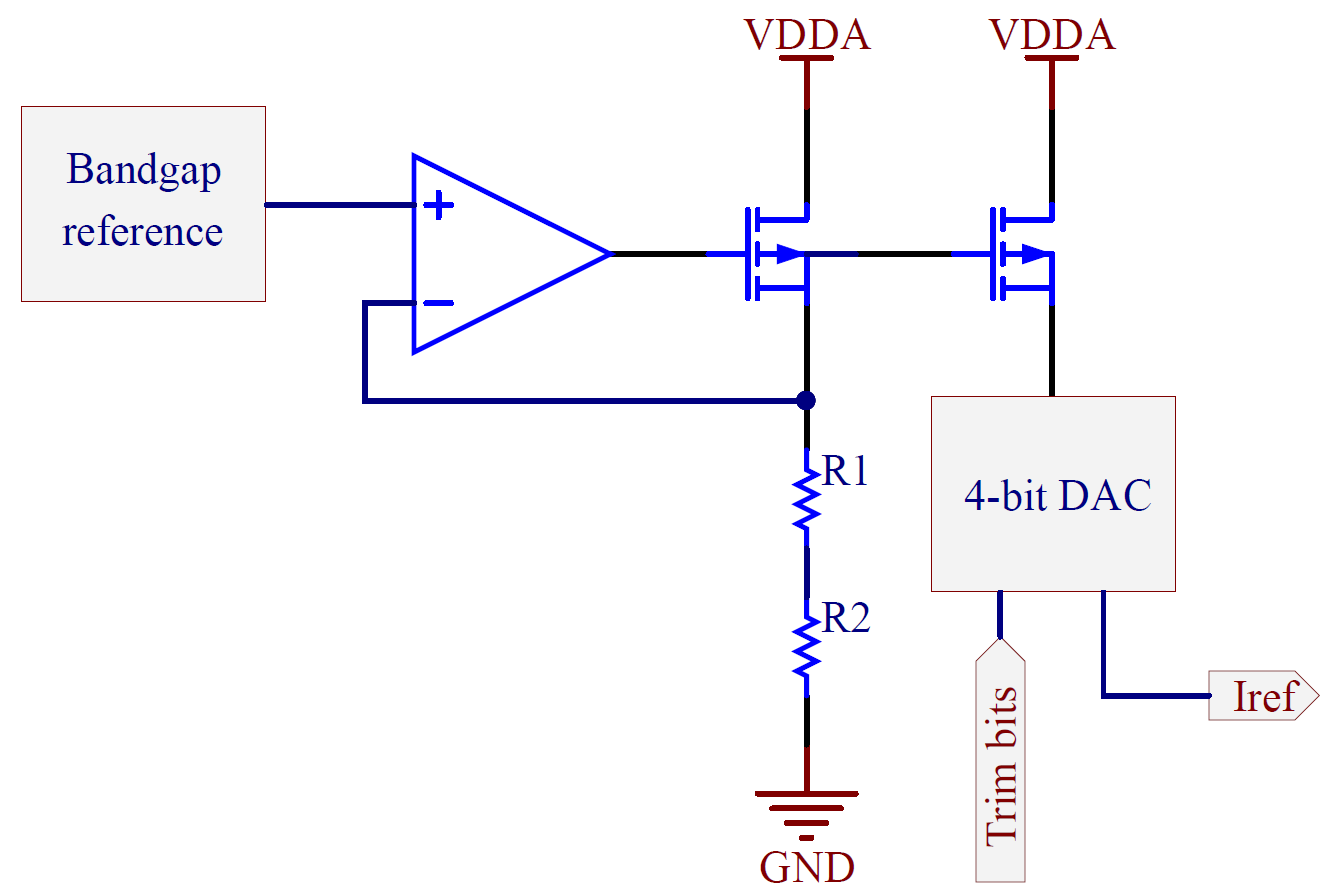}}
		\qquad
		\subfloat[\label{fig:iref:b}]{\includegraphics[width=.48\textwidth,origin=c]{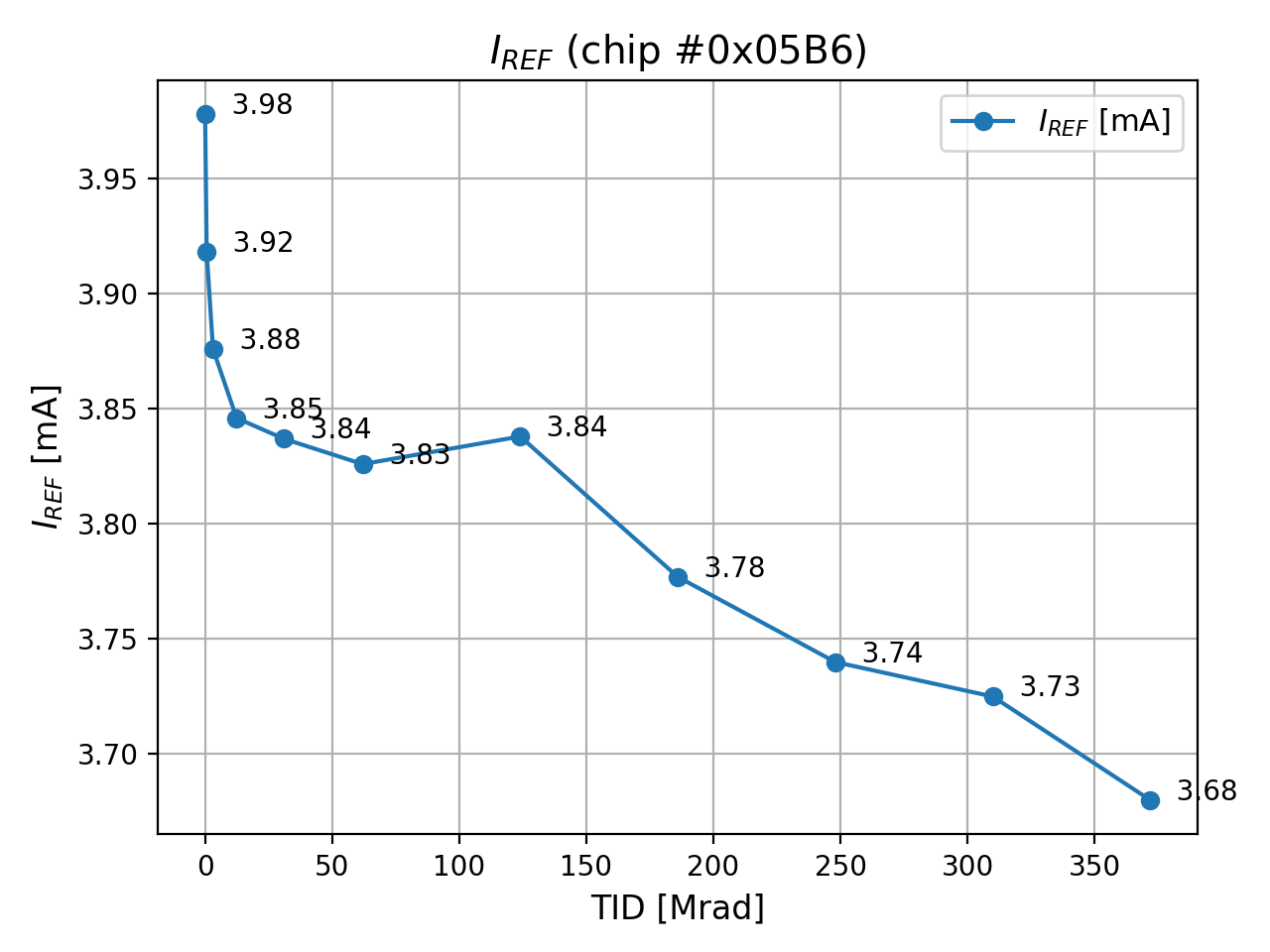}}\\[-2ex]
		\caption{\label{fig:iref}a) Schematic of the reference current source. b) $\rm I_{REF}$ as a function of the TID.}
	\end{figure}

	One of the crucial requirements for RD53A is a stable master reference current of \SI{4}{\micro\ampere}. As depicted in figure~\ref{fig:iref:a}, it is derived from a bandgap voltage source, followed by a voltage-to-current converter and a current mirror. Two resistors in series, $R1$ and $R2$, define the conversion factor. A 4-bit DAC can be used to trim the current and thereby compensate for process variations.
	
	During the campaign, $\rm I_{REF}$ was measured between the irradiation steps (figure~\ref{fig:iref:b}). The current degradation is significant for the first \SI{\ten}{Mrad} and becomes linear above $\sim 120 \rm Mrad$. After \SI{\sixhundred}{Mrad}, $\rm I_{REF}$ has decreased by $7.5\%$ of the initial value. 
	The reason for this behavior is still subject of investigation. One possible explanation lies in the design of the feedback resistors. In order to achieve high temperature stability, a combination of two different resistor implementations was used. The temperature coefficients of polysilicon and diffusion resistors partially compensate due to their opposite sign. The radiation sensitivity, especially of the diffusion resistor, is suspected to be responsible for the high reference current shift.
	
	This behavior was observed during multiple different X-ray TID campaigns within the collaboration as well. Consequently, RD53B will use external high precision metal film resistors instead, to reduce the probability of radiation-induced reference current changes.

	\subsection{CDR and CML output drivers}
	\label{sec:link}
\begin{figure}[htbp]
		\centering
		\subfloat[\label{fig:clock:a}]{\includegraphics[width=.485\textwidth,origin=c]{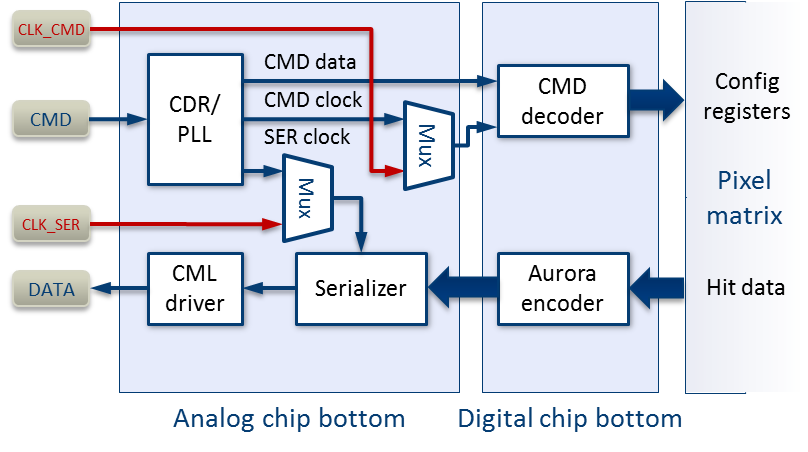}}
		\qquad
		\subfloat[\label{fig:clock:b}]{\includegraphics[width=.46\textwidth]{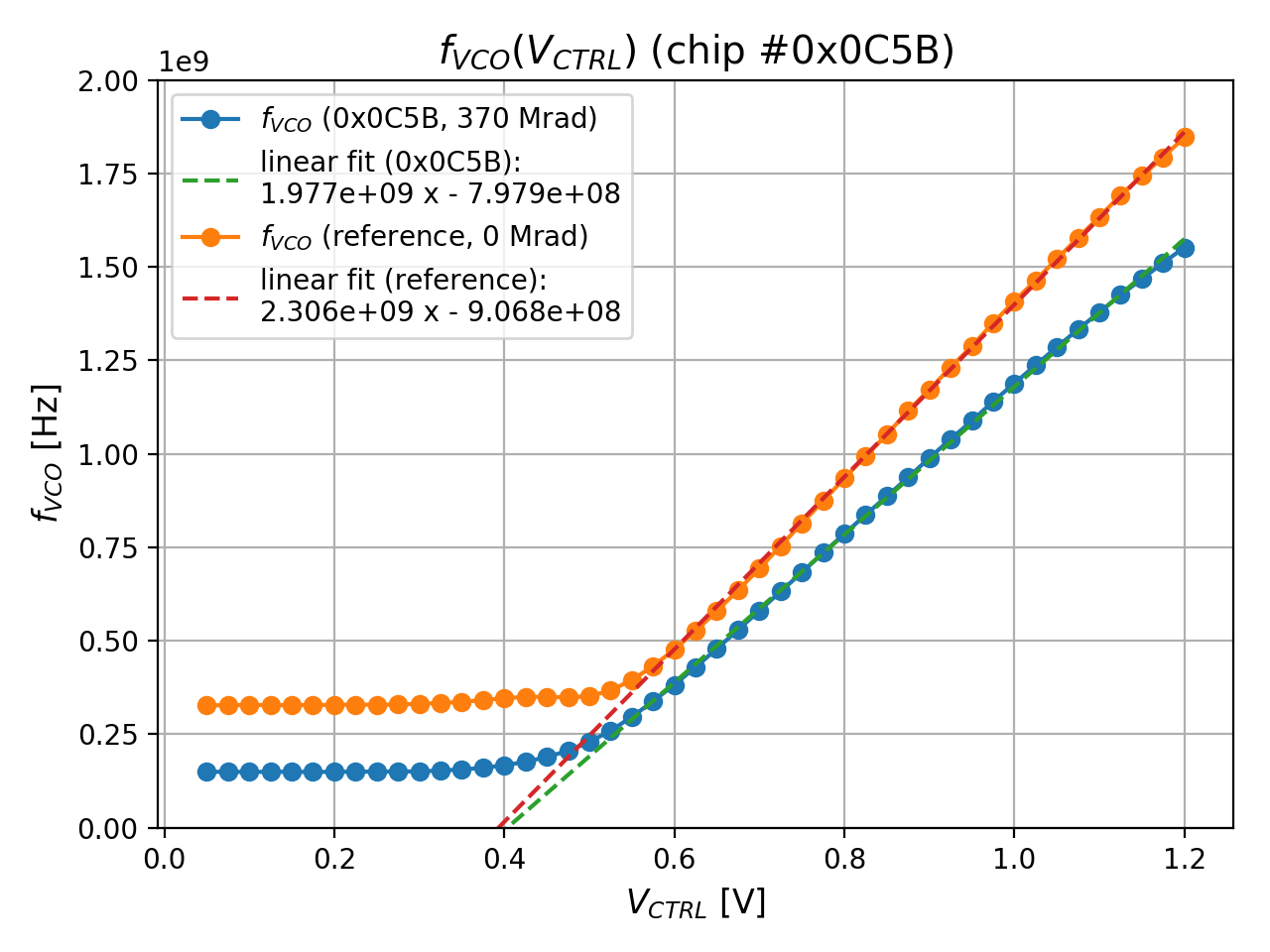}}\\[-2ex]
		\caption{\label{fig:clock}(a) Simplified clock generation and data flow blocks. (b) VCO gain curve measurement for a non-irradiated chip and after \SI{\sixhundred}{Mrad}.}
	\end{figure}

	In the default operation mode, the CDR block recovers the reference clock of nominally \SI{160}{\mega\hertz} from the received command data stream (figure \ref{fig:clock:a}). The CDR includes a Voltage Controllable Oscillator (VCO) and generates a high frequency clock, which is phase-locked to the recovered clock and used by the serializer to generate the $1.28\ \rm Gbit/s$ output data stream. By varying a control voltage ($\rm V_{CTRL}$), the CDR tunes the VCO frequency to a nominal value of $\rm f_{SER} = 8\times f_{CMD} = 1.28\ \rm GHz$. The VCO tuning range of a non-irradiated chip lies between $300\ \rm MHz$ and $1.9\ \rm GHz$. 
	
	\begin{figure}[htbp]
		\centering
		\includegraphics[width=.52\textwidth,origin=c]{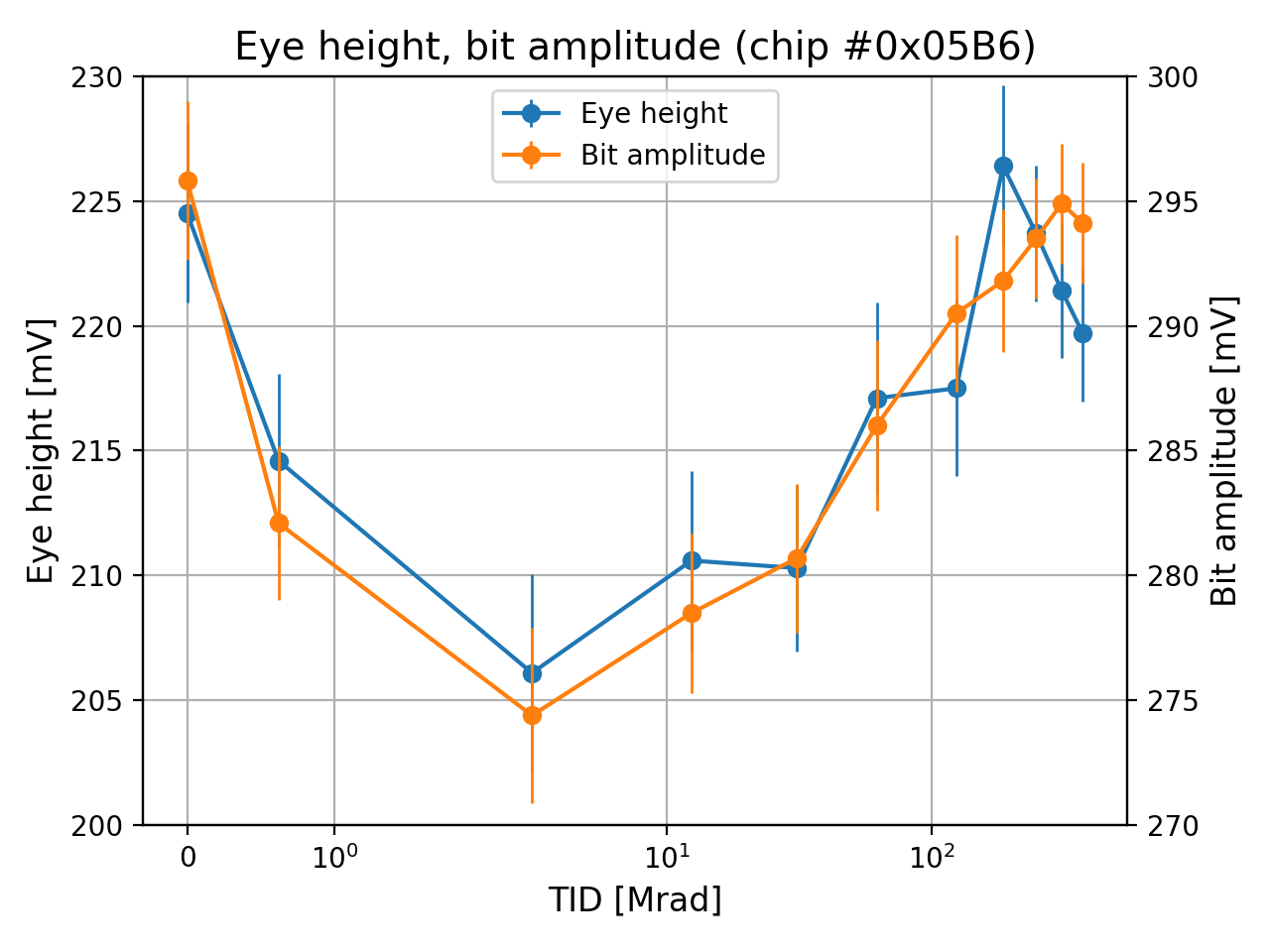}\\[-2ex]
		\caption{\label{fig:eye}Eye height and bit amplitude of the CML output driver.}
	\end{figure}
	
	The tuning range is expected to change after irradiation, as the propagation delays of the buffers in the VCO ring oscillator degrade. The VCO gain curve has been measured for a non-irradiated chip and after \SI{\sixhundred}{Mrad} with $\rm V_{CTRL}$ between \SI{25}{\milli\volt} and \SI{1.2}{\volt} (figure~\ref{fig:clock:b}). Compared to the non-irradiated sample, the VCO gain decreased slightly and the tuning range shifted towards lower frequencies by $\sim \rm 250\ MHz$. The nominal value of \SI{1.28}{\giga\hertz} can still be reached, thus the data link operation at \SI{1.28}{Gbit/s} is unaffected.	For RD53B, the VCO transistor sizes will be increased to improve the performance and radiation hardness.
	The cable drivers of RD53A are implemented as Current Mode Logic (CML) buffers with programmable pre-emphasis. The bit amplitude and the eye height (figure \ref{fig:eye}) have been measured as part of the measurement routines at each irradiation step. Both eye diagram characteristics fluctuated within $10 \%$ of their initial value during the irradiation.

	\subsection{Digital injection scans}
	\label{sec:shmoo}
\begin{figure}[htbp]
		\centering
		\subfloat[\label{fig:shmoo:a}]{\includegraphics[width=.46\textwidth,origin=c]{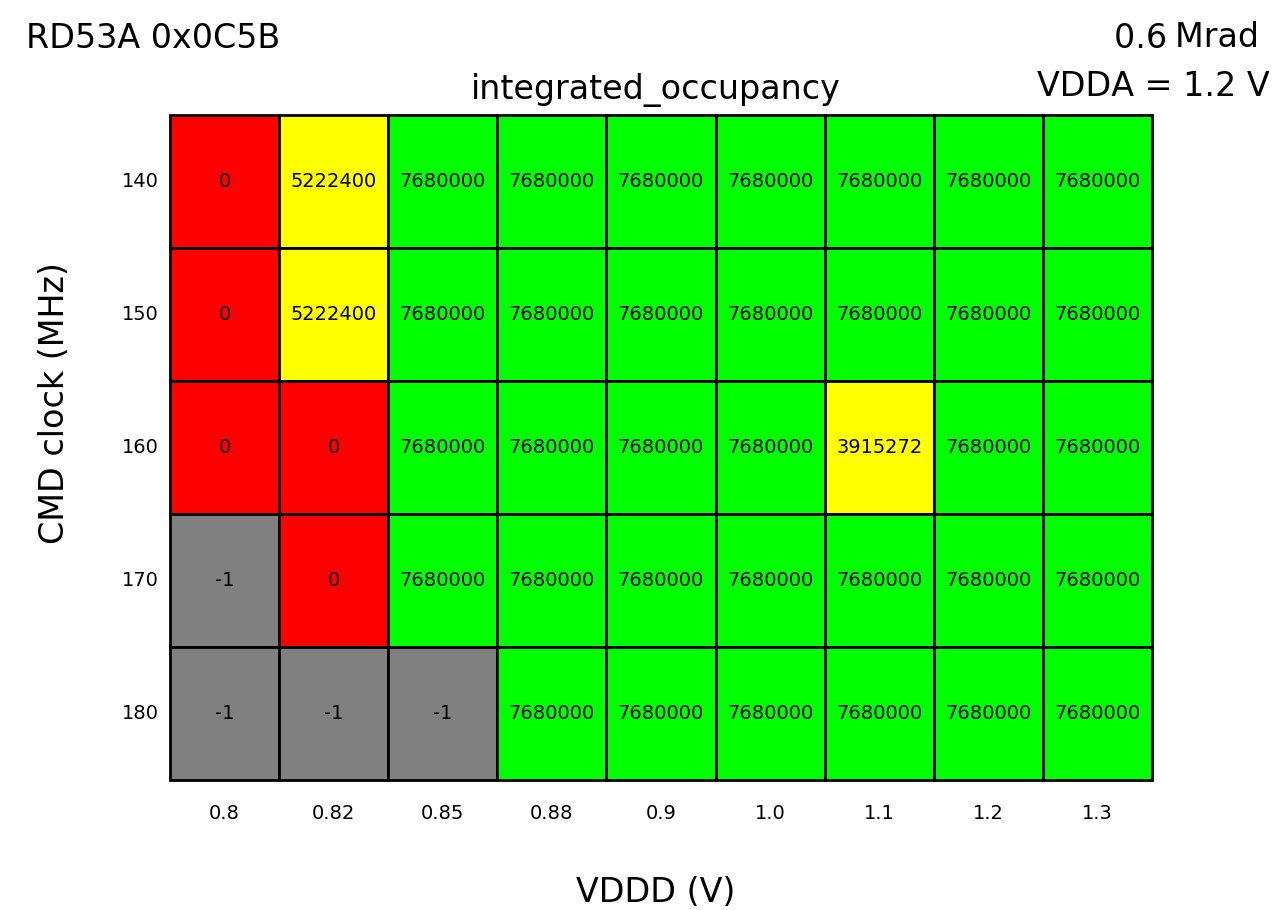}}\\[-2ex]
		\qquad
		\subfloat[\label{fig:shmoo:b}]{\includegraphics[width=.46\textwidth,origin=c]{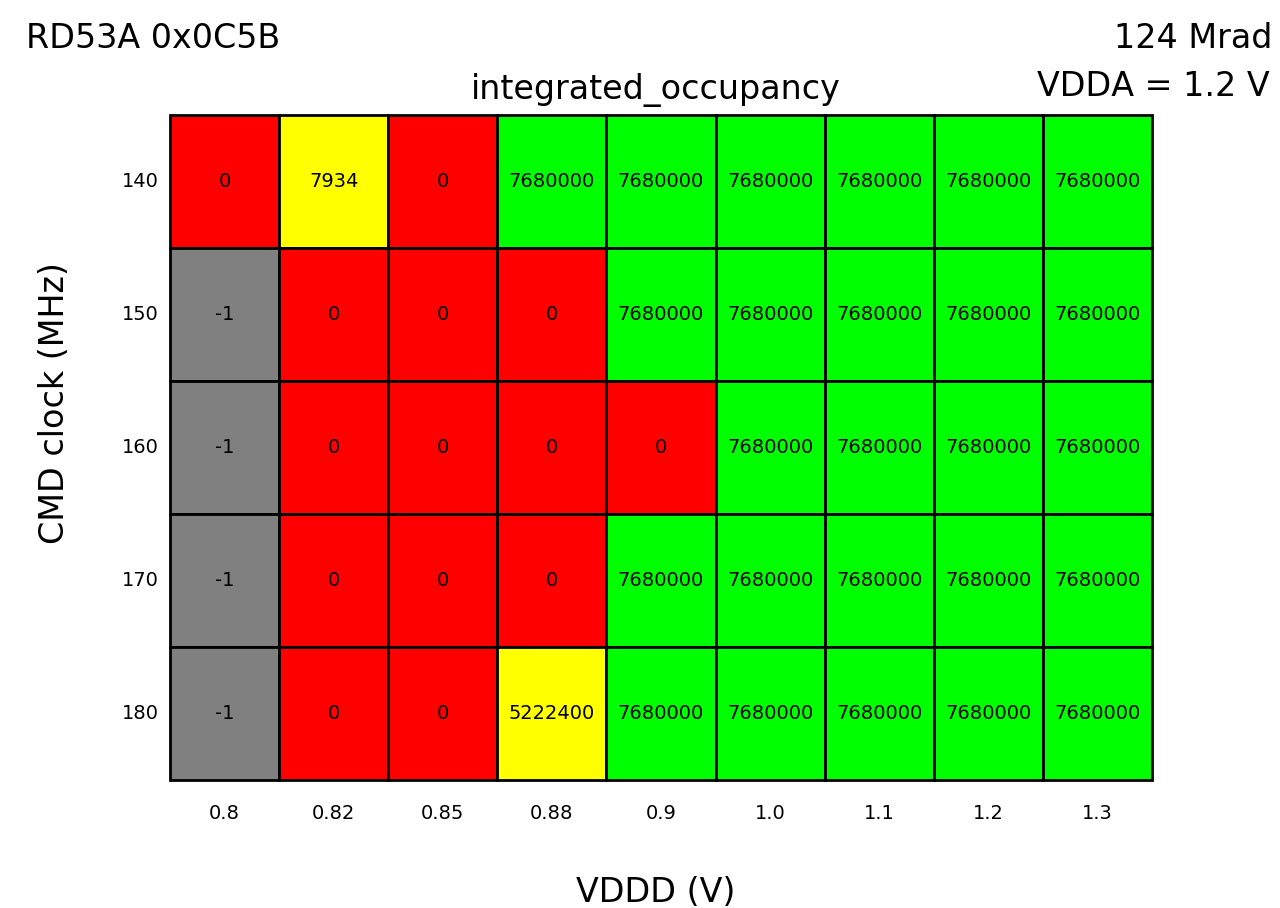}}
		\subfloat[\label{fig:shmoo:c}]{\includegraphics[width=.46\textwidth,origin=c]{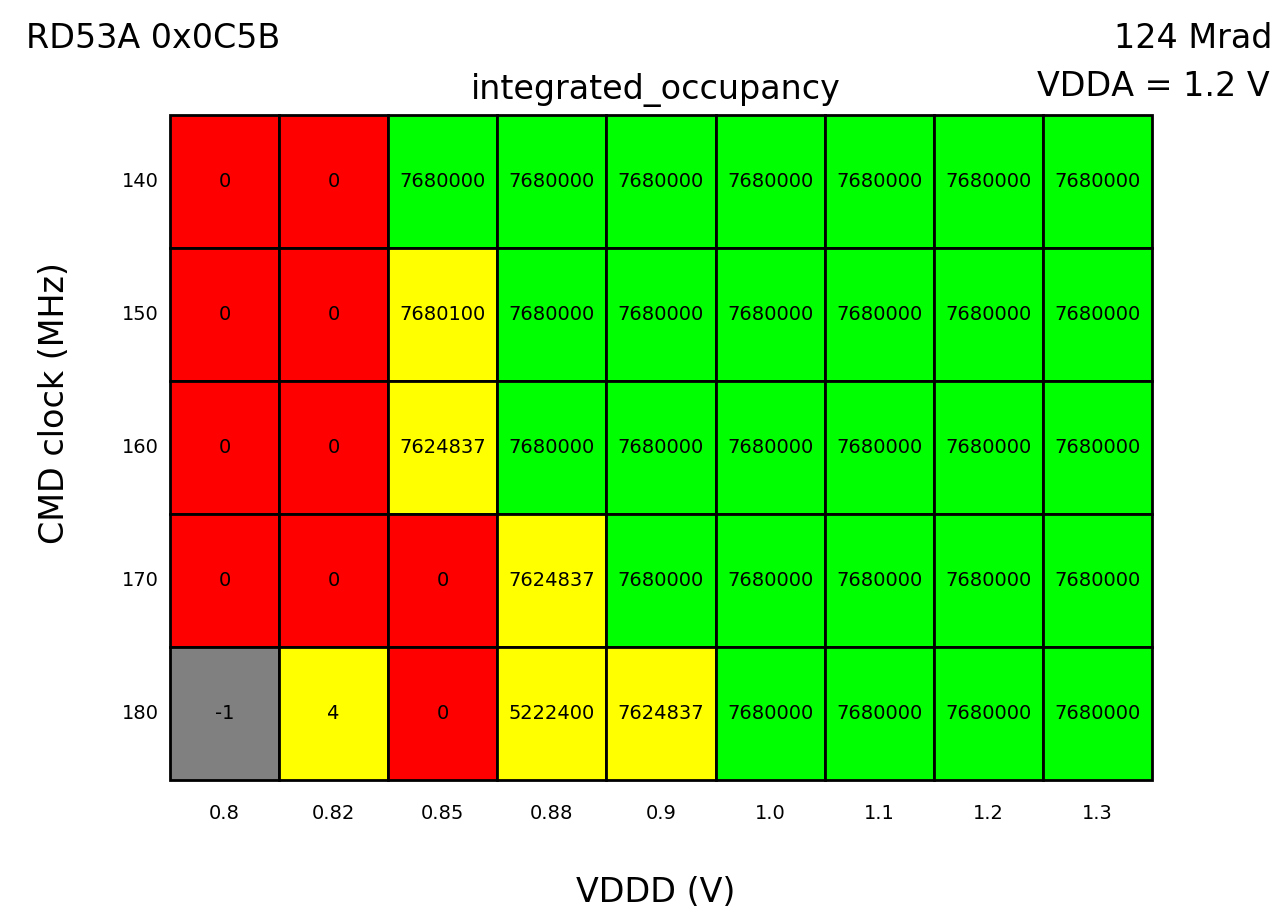}}\\[-2ex]
		\qquad
		\subfloat[\label{fig:shmoo:d}]{\includegraphics[width=.46\textwidth,origin=c]{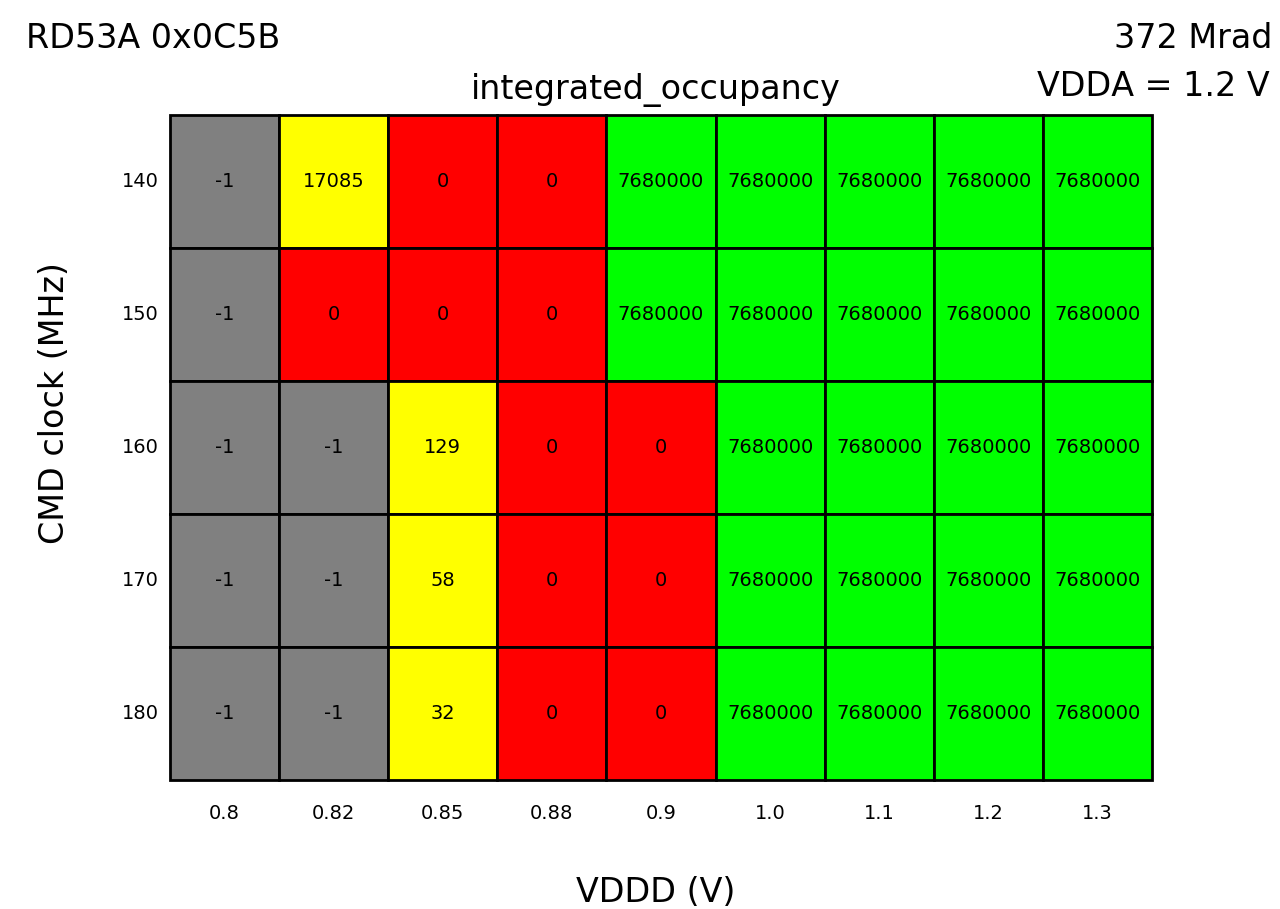}}
		\subfloat[\label{fig:shmoo:e}]{\includegraphics[width=.46\textwidth,origin=c]{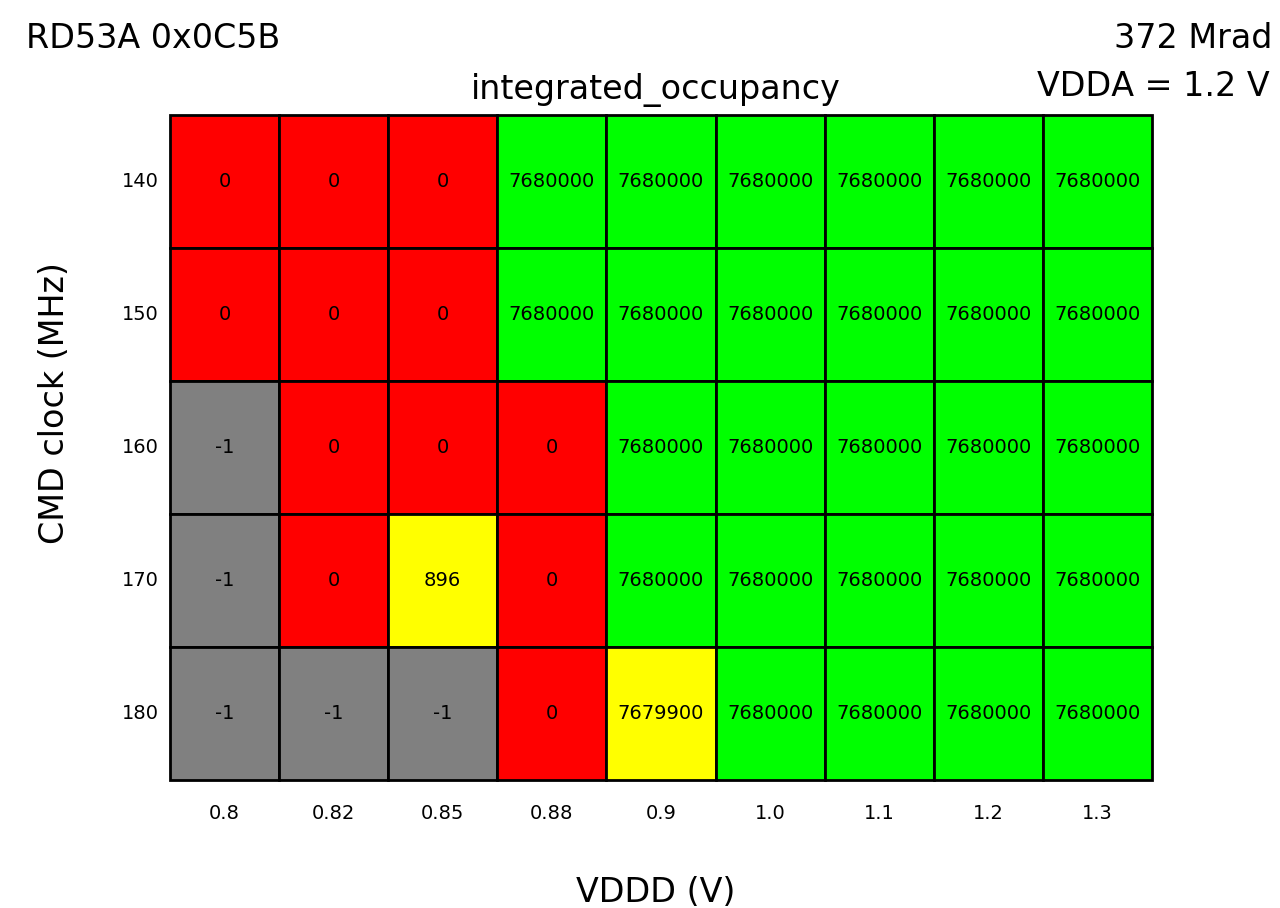}}\\[-2ex]
		\caption{\label{fig:shmoo}Shmoo plots for $\one$, $\twohundred$ and $\sixhundred\ \rm Mrad$. a), b) and d) show the POR behavior at $\rm V_{DDD}$, c) and e) represent POR at \SI{1.2}{\volt} (described in section~\ref{sec:por}).
		}
	\end{figure}

	A common method to visualize the response of integrated circuits to varying operating conditions is the \textit{shmoo plot} (figure~\ref{fig:shmoo}). Each field represents a combination of possible conditions, in terms of supply voltage and input clock frequency. The measured response is the result of a test procedure. In case of this RD53A irradiation campaign, the test is a digital injection scan and the parameters are: $0.8\ \rm V\le V_{DDD} \le 1.3\ \rm V$ and $140\ \rm MHz \le f_{CMD} \le 180\ \rm MHz$. The analog supply voltage $\rm V_{DDA}$, which does not affect the digital scans, is kept at a constant value of $1.2\ \rm V$ during all measurements. 
	
	For each combination of settings, the following measurement procedure is performed:
	\begin{itemize}[topsep=3pt,itemsep=-1ex,partopsep=1ex,parsep=1ex]
		\item power off the chip
		\item set $\rm V_{DDD}$ and $\rm f_{REF}$ to the new values
		\item power on, try to establish communication (up to 3 times)
		\item run a digital scan with 100 injections per pixel
	\end{itemize}
	
	The different colors in figure~\ref{fig:shmoo} represent the results of the digital scans, which inject 100 times per pixel into the digital hit processing chain. For the full matrix, this leads to an expected integrated occupancy of $7.68 \times 10^6$ hits. 
	For the extreme cases in the lower left corner of the plots, where the chip is operated at low voltages and high frequencies, no communication can be established - represented by gray colored entries.
	At higher voltages still below the nominal value of \SI{1.2}{\volt}, for some combinations the link can be established, but the injections are either not registered at all or they are not fully processed. Those cases are represented by red and yellow entries, respectively. The remaining green entries depict the combinations, which yield the expected occupancy.
					
	\subsection{Power On Reset}
	\label{sec:por}
	The digital logic of RD53A is designed to operate with $\rm V_{DDD}$ down to $0.9\ \rm V$ at the default reference clock frequency of \SI{160}{\mega\hertz}.
	However, after \SI{\twohundred}{Mrad}, the chip could not be reset reliably at this voltage.
	The Power On Reset (POR) was investigated as the potential source of this behavior.
	A different POR method was introduced with an additional scan. Compared to the procedure described in section~\ref{sec:shmoo}, the initial $\rm V_{DDD}$ value was fixed to \SI{1.2}{\volt} during the POR and afterwards decreased to the desired value. This ensures that the chip reset is always performed at the same voltage.
	As shown in figure~\ref{fig:shmoo:c} and \ref{fig:shmoo:e}, the operating margin is increased, especially for \SI{\twohundred}{Mrad}. 
	
\begin{figure}[htbp]
		\centering
		\includegraphics[width=.6\textwidth,origin=c]{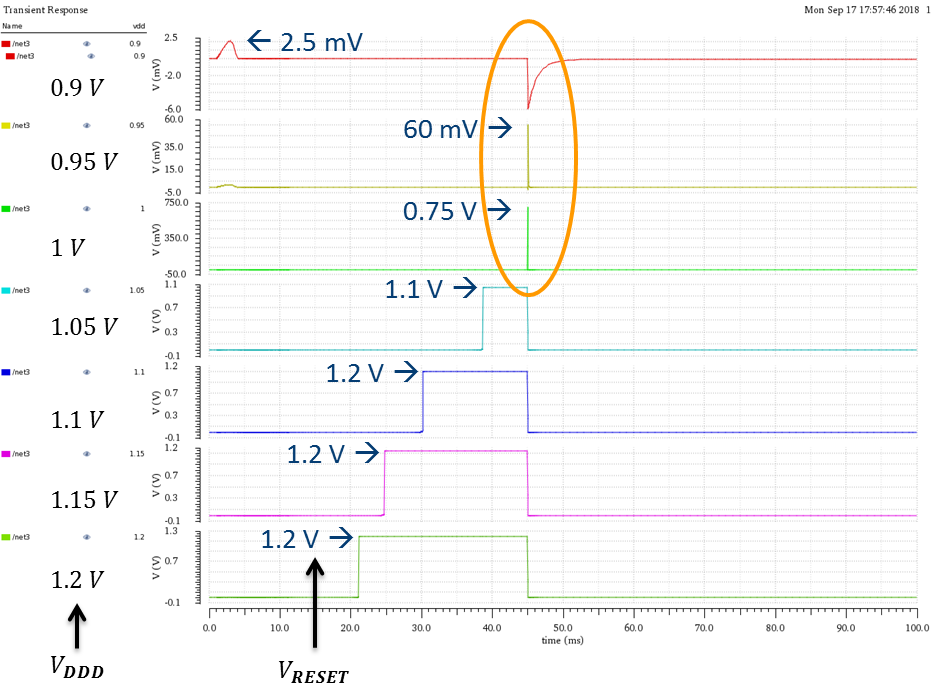}\\[-1ex]
		\caption{\label{fig:por}Simulation of the power-on reset pulse width and amplitude with varying $\rm V_{DDD}$.}
	\end{figure}

	The simulation at $-5\ ^\circ C$ and variable $\rm V_{DDD}$ (figure \ref{fig:por}) shows that the internal reset signal width decreases to a narrow pulse with reduced amplitude for $\rm V_{DDD} \le 1\ \rm V$, which is insufficient to perform a successful chip reset.
	The POR circuit was designed using the analog corner, assuming a minimum  $\rm V_{DDD}$ of \SI{1.05}{\volt}. Since the operation voltage will be larger than \SI{1.1}{\volt} during the lifetime of the readout chip, this limitation is not critical.
	
	\subsection{Command clock phase shift}
	\label{sec:phase}
	In CDR-bypass mode, the phase relation between the external command clock and the command data is critical, as it is not anymore controlled by the CDR. Instead, the command data signal is sampled at the rising edge of the command clock. If this mode is used, the phase needs to be set correctly. 
	
\begin{figure}[htbp]
		\subfloat[]{\includegraphics[width=.48\textwidth,origin=c]{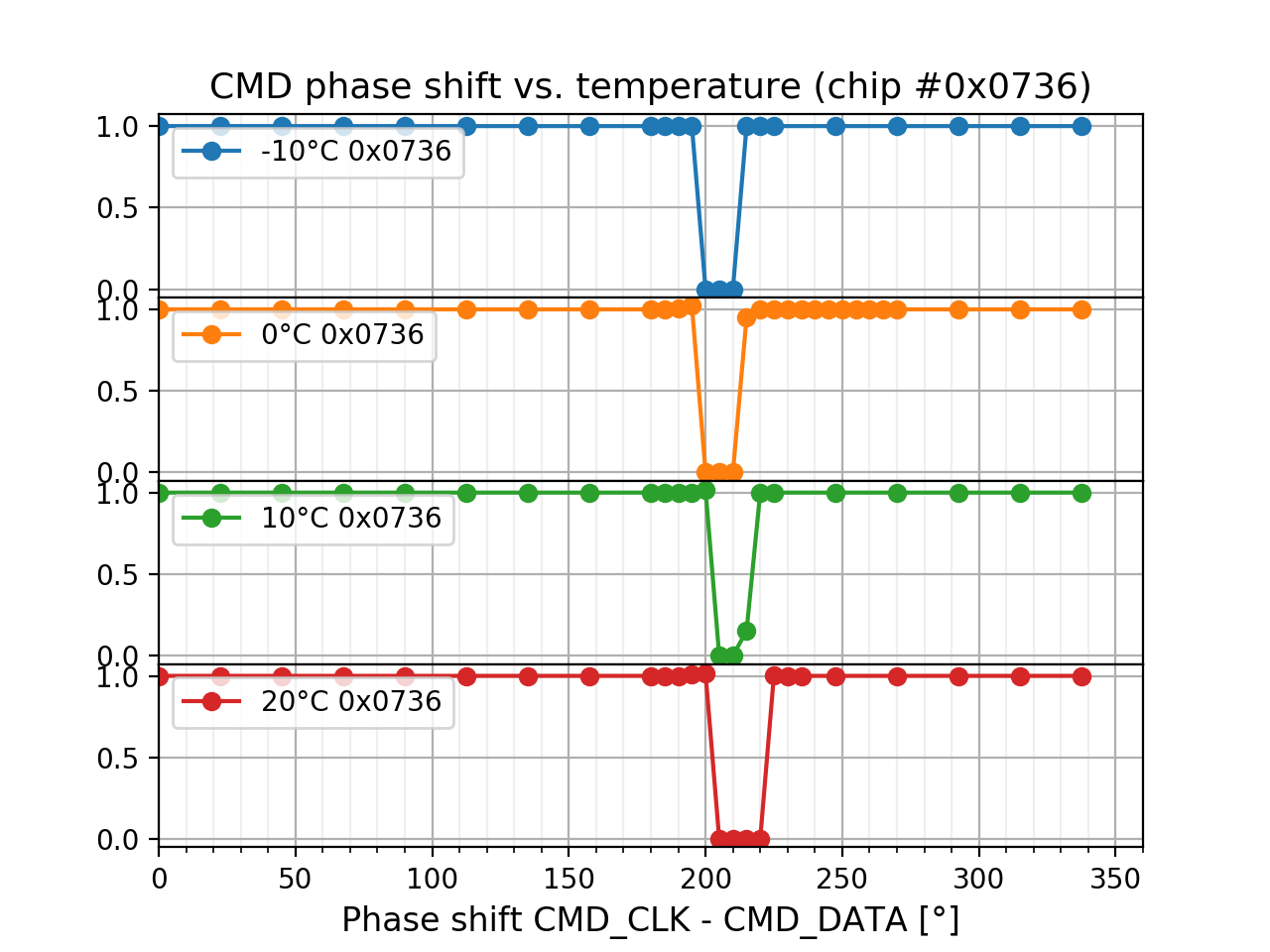}}
		\qquad
		\subfloat[]{\includegraphics[width=.48\textwidth,origin=c]{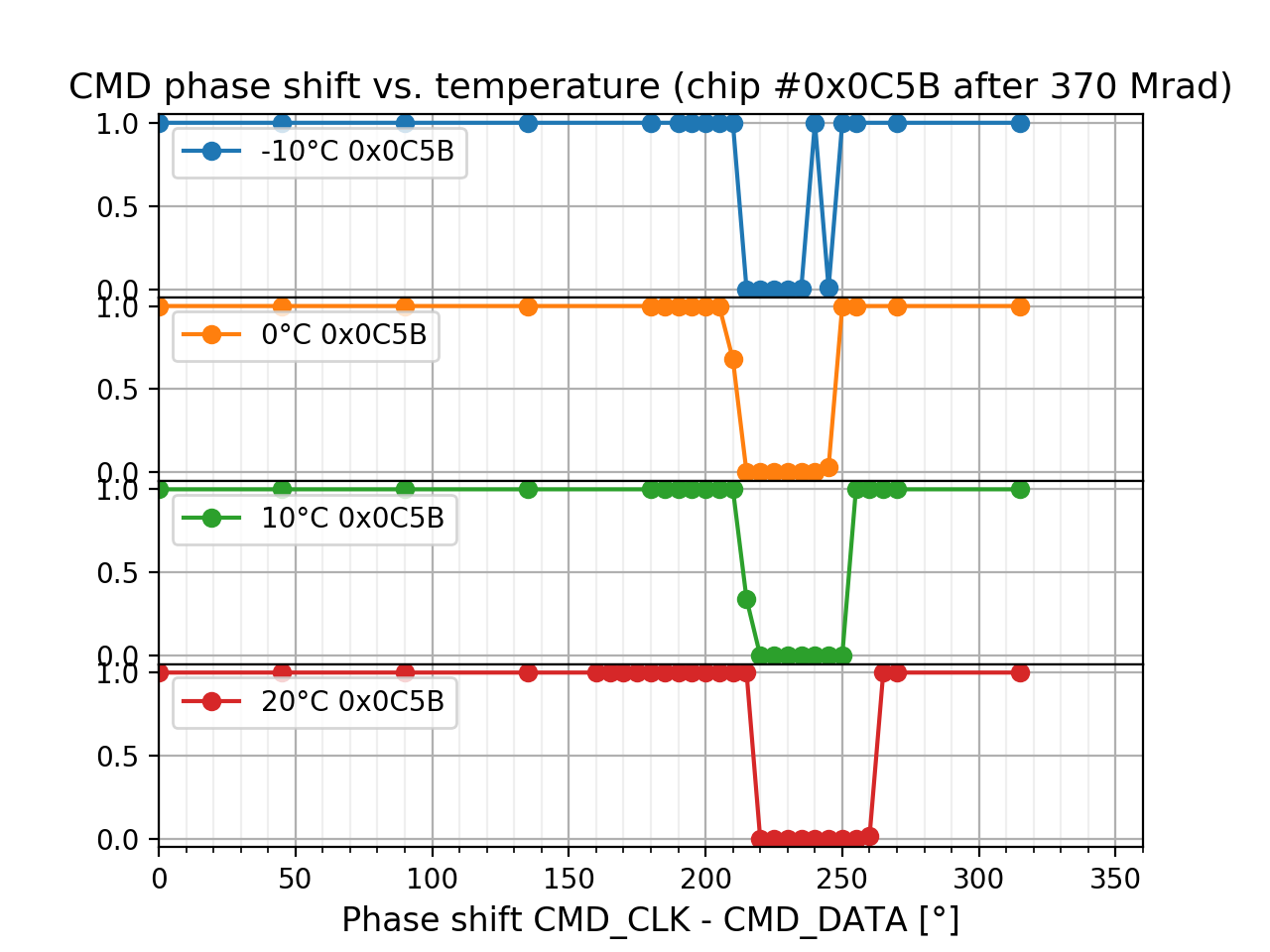}}\\[-1ex]
		\caption{\label{fig:cmd_phase}Phase margin of the command data signal in CDR-bypass mode}
	\end{figure}

	Measurements with an external dual-channel clock generator were performed at different temperatures after the campaign.
	They revealed, that the minimum hold time (distance between data transition and the next clock edge) increases by $15^\circ$ after a TID of \SI{\sixhundred}{Mrad}, compared to a reference chip (figure \ref{fig:cmd_phase}). At a temperature of $-20^\circ C$, the critical phase region width increases from $20^\circ$ at \SI{\ten}{Mrad} to $45^\circ$ at \SI{\sixhundred}{Mrad}. The temperature dependence is $\sim0.16^\circ/^\circ \rm C = 2.9\ \rm ps/^\circ \rm C$ for both samples.
	Therefore, even after \SI{\sixhundred}{Mrad}, the timing is uncritical, as long as the readout system synchronizes the command data with the falling edge of the clock signal.
	
	\section{Conclusion and outlook}
	\label{sec:conclusion}
	RD53A has been irradiated to \SI{\sixhundred}{Mrad}.
	No significant degradation of the CML driver and the VCO tuning range has been observed. 
	The POR circuit is not reliable at \rm $\rm V_{DDD}<1\ \rm V$ after $\sim$\SI{120}{Mrad}. However, this behavior is not supposed to cause issues in the default operation conditions of the experiment.
	The reference current showed significant radiation sensitivity and has decreased by $7.5\%$. This will be mitigated in RD53B, by the usage of external precision resistors.
	
	Nevertheless, further measurements are required to gain a better understanding of radiation damages under different operating conditions, in order to find mitigation strategies for RD53B and the final pixel readout chip. For example, a low dose rate irradiation with a $\rm ^{85}Kr$ source is in progress, and a low dose X-ray campaign is starting soon.
	
	\acknowledgments
	This project has received funding from 
	the German BMBF under grant No.\ 05H15PDCA9,
	European Union's Horizon 2020 Research and Innovation programme under grant agreements No.\ 654168 (AIDA-2020) 
	and 675587 (Marie Sk\l{}odowska-Curie ITN STREAM).

\end{document}